# Multicomponent Isotope Engineering of Phonon Polaritons in Hexagonal Boron Nitride

**Nihar Ranjan Sahoo[1], Byung-Il Noh[2], Brandon Stephanie[1], Hossein Zandipour[1], Eli Janzen[3], James H. Edgar[3], Siyuan Dai[2,†] and Thomas G. Folland [1,*]**

[1]Department of Physics and Astronomy, University of Iowa, Iowa City, Iowa 52245, USA

[2] Materials Research and Education Center, Department of Mechanical Engineering, Auburn University, Auburn, AL, 36849, USA

[3] Tim Taylor Department of Chemical Engineering, Kansas State University, Manhattan, KS 66506, USA

[*] thomas-folland@uiowa.edu

[†] sdai@auburn.edu

**Abstract**

Isotopic enrichment provides a non-invasive means of tuning the vibrational and optical properties of van der Waals crystals (vdW) without altering their lattice structure. We investigate two doubly enriched hexagonal boron nitride (hBN) compositions, $h^{10}B^{15}N$ and $h^{11}B^{15}N$, to determine how isotope substitution on both sublattices modifies mid-infrared hyperbolic phonon polaritons (HPhPs). Polarization-resolved room temperature and cryogenic far-field reflectance measurements reveal optical-phonon blue shifts of approximately 35 $cm^{-1}$ in $h^{10}B^{15}N$ relative $h^{11}B^{15}N$, translating both Reststrahlen bands (RBs) and their polariton dispersions. For the upper RB, the combined accessible range expands from 1360-1614 $cm^{-1}$ in natural hBN to 1356-1640 $cm^{-1}$ across the B- enriched $^{14}N$ crystals and 1339-1640 $cm^{-1}$ when the $^{15}N$ enriched crystals are included. Near-field imaging confirms deeply sub-diffractional propagation and shows that both crystals exhibit lower polariton damping and longer lifetimes than natural hBN, with performance comparable to monoisotopically enriched crystals. Further, we compare the conventional reduced-mass description with a phenomenological two-component inverse-mass model that captures the isotope dependence of the phonon frequencies and LO–TO coupling while resolving the contributions of the individual sublattices. These results establish multicomponent isotope engineering as a materials design strategy for positioning the HPhPs spectral range while preserving low-loss propagation in polar vdW crystals.

## Introduction

Infrared (IR) nanophotonics focuses on manipulating light at deeply sub-wavelength scales, enabling advances in molecular sensing, thermal emission, communications, and imaging.[1–6] A central approach is to exploit polaritons, hybrid light–matter coupling to confine and guide IR electromagnetic fields far below the diffraction limit.[1,7,8] In particular, HPhPs supported by polar vdW crystals have emerged as a promising platform for low-loss IR nanophotonics.[8–12] Owing to the long lifetimes of optical phonons, PhPs resonances can exhibit significantly lower optical losses and higher quality factors than plasmonic modes at comparable frequencies.[13–15] These deeply sub-diffractional modes span the critical mid-IR "molecular fingerprint" region (~6–12 µm), enabling strong light–matter interactions with vibrational absorption for chemical identification and sensing.[16–18] Hexagonal boron nitride; a naturally hyperbolic vdW crystal, supports PhPs across two RBs (approximately 760–825 $cm^{-1}$ and 1360–1610 $cm^{-1}$) that overlap many molecular vibrational signatures, suitable for IR nanophotonic applications.[9,14,17]

Despite the advantages, existing phonon-polaritonic materials face two key limitations: limited spectral tunability and residual optical losses. The operating frequencies of HPhPs are fixed by the host crystal's phonon modes, which means one cannot easily adjust polariton resonances to an arbitrary IR frequency.[19] For example, natural hBN and natural cubic BN (cBN) collectively cover a broad mid-IR range (Upper RB: 1050-1614 $cm^{-1}$), the upper cBN RB ends near 1305 $cm^{-1}$ and the upper hBN band begins near 1364 $cm^{-1}$, leaving a ~60 $cm^{-1}$ (~7.7 µm) gap between the two materials' bands.[19,20] In general, accessing different target frequencies across the mid-IR has required selecting different polar crystals, such as hBN, alpha phase molybdenum trioxide ($\alpha$-$MoO_3$), alpha phase vanadium pentoxide ($\alpha$-$V_2O_5$) or beta phase gallium oxide ($\beta$-$Ga_2O_3$) whose optical phonons define distinct and intrinsically fixed RBs,[14,21–25] or employing active modulation[26–29], underscoring the lack of intrinsic tunability within any single vdW crystal. At the same time, optical losses, though lower than in plasmonic systems, remain significant due to lattice disorder and anharmonic phonon decay. In naturally abundant hBN, the natural mixture of boron isotopes (~20% $^{10}B$ and 80% $^{11}B$) is a built-in source of mass disorder that scatters optical phonons.[30] This isotopic disorder broadens optical phonon lines and limits phonon-polariton lifetimes, imposing a fundamental performance limit even in high quality hBN devices.[30,31] These challenges motivate new materials engineering strategies to broaden spectral tunability while further reducing optical damping in phonon-polaritonic systems.

Isotope engineering of polar vdW crystals has emerged as a viable approach to address both limitations. Isotopic enrichment alters a material's phonon frequencies by changing the atomic masses, and at the same time it can reduce phonon scattering by reducing isotope mass disorder.[23,31–34] In hBN, enriching the crystal in a single isotope (monoisotopic hBN) was shown to alter its phonon-polariton response. Nearly monoisotopic hBN (containing > 99% $^{10}B$ or $^{11}B$ in the boron sublattice) exhibits substantial shifts in optical phonon frequency on the order of tens of $cm^{-1}$ compared to natural hBN, along with narrower phonon linewidths and longer phonon lifetimes.[31,35–38] For example, replacing $^{11}B$ with the lighter $^{10}B$ in hBN blue shifts the TO phonon frequency by ~30–40 $cm^{-1}$, and removing boron isotope disorder yields much sharper vibrational linewidths.[31] Further, hyperbolic PhPs in isotopically pure hBN propagate with lower damping and longer lifetime,[35] experiments report approximately two-fold increase Q factors, along with longer polariton propagation lengths, in ~99% $^{10}B$- or $^{11}B$-enriched hBN relative to natural hBN.[31,37] By removing the principal source of lattice disorder, isotope enrichment establishes a clear route to improving PhPs performance, as well as expanding the accessible spectral range for high quality phonon polariton resonances.

However, all prior studies on polaritonics have focused on boron-enriched crystals with natural abundance nitrogen in the hBN lattice, leaving the effects of multicomponent isotope substitution largely unexplored. In principle, simultaneously tailoring the isotopic composition of both sublattices could provide additional control over phonon frequencies and damping beyond what single-element enrichment offers. Recently nitrogen-15 isotope enriched ($^{15}N$) hBN was synthesized using a two-step growth approach for high-quality $h^{10}B^{15}N$ and $h^{11}B^{15}N$ crystals and established their isotope-dependent structural, vibrational, and electronic properties through Raman spectroscopy, photoluminescence, and first-principles phonon calculations, which provide an important materials and vibrational foundation for the present study.[39–41] However, the far-field infrared dielectric response and near-field HPhPs behavior of these doubly isotope enriched crystals, including their ultralow-loss polariton dispersion, have not been systematically investigated.

Here, we compare the polariton response of two doubly enriched hBN crystals, $h^{10}B^{15}N$ and $h^{11}B^{15}N$. By varying the boron isotope (lighter $^{10}B$ vs $^{11}B$) while maintaining an nitrogen enriched $^{15}N$ sublattice, allowing us to isolate the impact of boron mass on polariton properties under conditions of minimal isotopic disorder. We combine polarization-resolved cryogenic far-field and near-field infrared measurements to characterize each isotopically engineered crystal, to determine isotope-dependent phonon shifts and polariton propagation. Across the isotopically engineered hBN compositions examined here and in prior work, isotope substitution provides a phonon-frequency tuning range of approximately 35-55 $cm^{-1}$ for transverse optic phonon frequency across both RBs. Here, both doubly enriched crystals exhibit longer propagation lengths and higher Figure of Merit (FOM) than natural hBN, with $h^{10}B^{15}N$ providing a higher FOM across most of the upper RB than $h^{11}B^{15}N$. Our findings therefore establish multicomponent isotope engineering as a practical route for quantitatively tuning polariton frequencies and reducing optical loss in hBN, providing a useful additional degree of freedom for mid-IR nanophotonic design. To interpret the frequency shifts, we compare the conventional reduced mass description with a two-component inverse mass model fitted to the transverse optic (TO) and longitudinal optic (LO) frequencies. Applying the same analysis to $\alpha$-$MoO_3$ shows relatively balanced B and N contributions in hBN, but a strongly oxygen dominated response in $\alpha$-$MoO_3$ that the reduced mass description does not capture. Together, these results shown how isotope composition can shift the polariton spectral range while retaining low-loss propagation.

## Results and Discussion

We first examined isotope substitution effects on hBN's far-field reflectance. Fig. 1a illustrates the micro-FTIR reflectance setup (see Methods), and corresponding Fig. 1(b-c) shows reflectance spectra from $h^{10}B^{15}N$ and $h^{11}B^{15}N$ flakes. In the far-field measurements, a custom aperture plate placed over the high numerical aperture objective restricts the illumination to a selected angular sector, thereby minimizing polarization mixing and providing a well-defined *p*-polarized excitation geometry.[42,43] (Fig. 1a Inset) This approach enhances sensitivity to the electric-field component normal to the hBN basal plane, enabling more reliable resolution of the out-of-plane phonon response and its isotope dependence. The reflectance spectra exhibit the characteristic broad RBs of hBN (arising from TO and LO phonons) with Fabry–Pérot interference fringes due to the finite flake thickness. Relative to $h^{11}B^{15}N$, $h^{10}B^{15}N$ exhibits clear blue shifts of both RBs, consistent in magnitude with the shifts previously observed when $^{11}B$ is replaced by $^{10}B$ in $^{14}N$ based hBN. Specifically, the TO(LO) frequencies shifts by 24 (20) $cm^{-1}$ in the lower

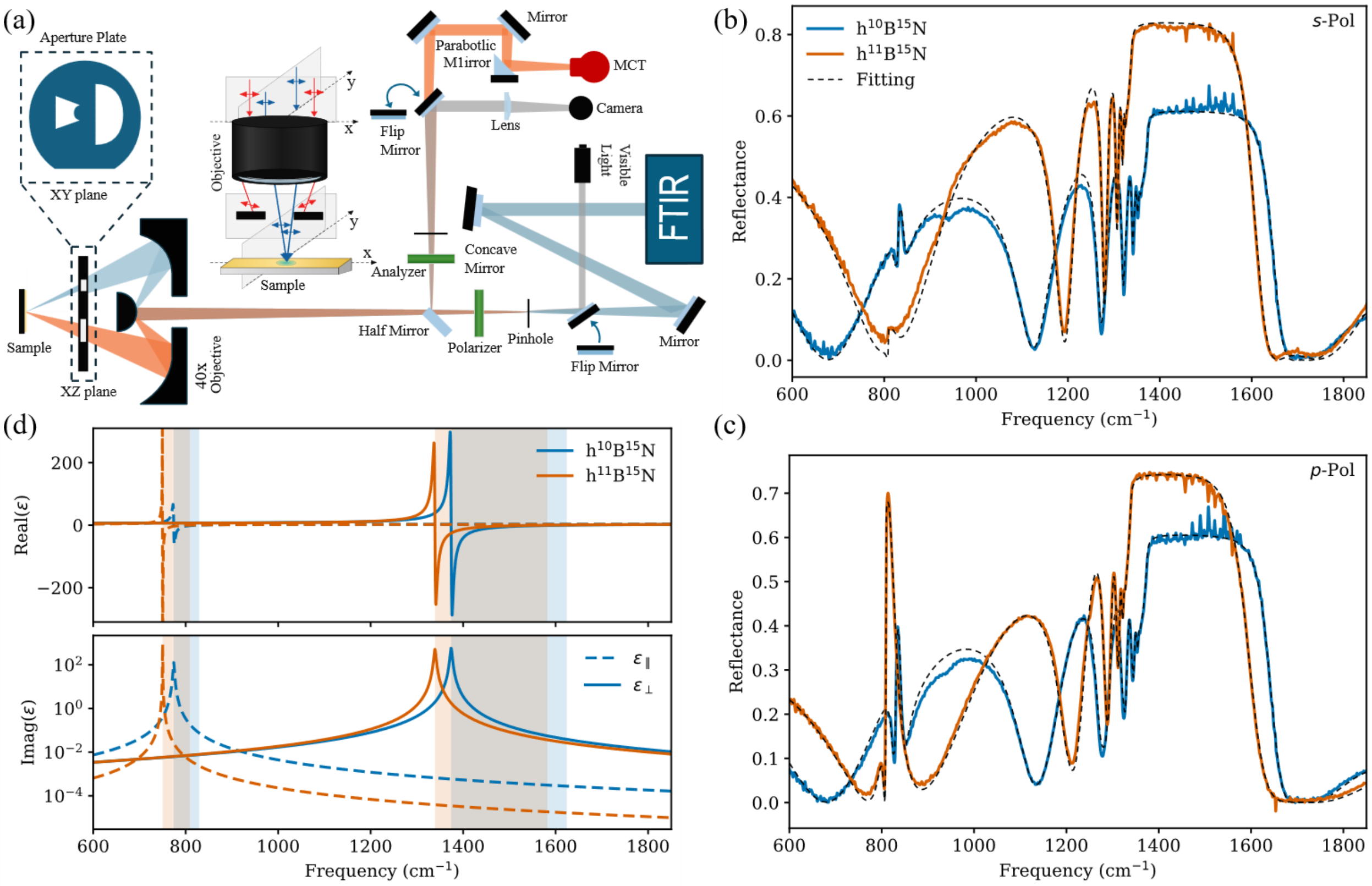

*Figure 1 : Polarization-resolved far-field response of isotope engineered hBN. (a) schematic of the custom built micro FTIR system, Inset shows the aperture plate used to restrict the illumination angle sector, (b-c) optical response of isotopically engineered hBN crystals, $h^{10}B^{15}N$ and $h^{11}B^{15}N$ for both s- & p-polarized light. Colored solid curves are the measured spectra, and black dashed curves are the corresponding fitting to extract the dielectric parameters. (d) Real and imaginary parts of the extracted permittivity tensor components ($\varepsilon_{\perp}$ and $\varepsilon_{\parallel}$) plotted as a function of frequency, showing blue shifts in both RBs in $h^{10}B^{15}N$ relative to $h^{11}B^{15}N$, due to isotope substitution. The shaded regions indicate the relevant RBs for $h^{10}B^{15}N$ (blue) and $h^{11}B^{15}N$ (orange), respectively.*

RB, and 35 (42) $cm^{-1}$ in the upper RB, compared with corresponding shifts of 30 (31) and 34 (41) $cm^{-1}$ respectively, between $h^{11}B^{15}N$, $h^{10}B^{15}N$. This shift is a consequence of the lighter average atomic mass in $h^{10}B^{15}N$, which increases the phonon frequencies (approximately following $\omega \propto \mu^{-1/2}$ where μ is the reduced mass). Notably, isotope substitution changes the atomic mass without altering the chemical composition or crystal symmetry of hBN; therefore, the phonon-frequency shifts are attributed primarily to the mass difference, although isotopic purity and growth-related disorder may also influence the phonon linewidths and polariton losses.[30,36] In addition, the fitted dielectric tensor from the polarized reflectance measurements confirm that both isotopically enriched crystals retain strong hyperbolic response and uniaxial anisotropy. As expected for hBN, when we measured reflectance for *s*- and *p*-polarized illumination in the lower RB (~780–830 $cm^{-1}$, where $\mathrm{Re}\,(\varepsilon_{\parallel}) < 0$), *p*-polarized illumination (having a strong out-of-plane electric field component due to the high NA objective) couples efficiently to the out-of-plane phonon mode, producing a reflectance peak, (Fig. 1c) whereas *s*-polarization (in-plane component) shows a suppressed response. (Fig S1 of Supporting Information S1) In the upper RB (~1360–1610 $cm^{-1}$, where $\mathrm{Re}(\varepsilon_{\perp}) < 0$), by contrast, both polarizations excite the in-plane phonon mode and yield nearly identical spectra.

Using a single oscillator dielectric fit to the reflectance of each sample, we obtained the complex permittivity spectra (Fig. 1d). Bulk, several micron free standing crystals were used so that the measured reflectance could be interpreted as a thin film supported in air. This geometry removes substrate contributions and allows for more accurate measurement of LO phonon energies, as well as the high frequency permittivity. This reduces correlations among the fitted parameters and enables a more direct extraction of intrinsic dielectric tensor than thin film methods used in some prior work. It also reduces the possibilities of inhomogeneities in the sample which can occur when thicker samples are used. The fitted dielectric functions for $h^{10}B^{15}N$ and $h^{11}B^{15}N$ show the expected two RBs with blue-shifted TO and LO phonon frequencies in the $h^{10}B^{15}N$ sample. (see Table 1) The in-plane TO phonon frequency is 1374 $cm^{-1}$ in $h^{10}B^{15}N$, compared with 1339 $cm^{-1}$ in $h^{11}B^{15}N$, corresponding to a 35 $cm^{-1}$ blue shift. Similarly large shifts are observed in the out-of-plane phonon frequencies. In addition to the room-temperature studies, we characterized the bulk dielectric response of free-standing $h^{11}B^{15}N$ crystal from 50K to 300K using variable-temperature micro-FTIR. (see Methods) The optical phonon frequencies and damping that were independently obtained at 300 K agree with the room-temperature measurements discussed above. As the temperature decreases, the TO phonon frequency increases and the linewidth narrows, following trends consistent with those previously reported for $h^{11}B^{15}N$ [30,35] while retaining the spectral shift associated with $^{15}N$ substitution. (Table S1 & Fig S2 of Supporting Information S2)

| Sample | $\varepsilon_\infty^\parallel$ | $\omega_{TO}^\parallel$ ($cm^{-1}$) | $\omega_{LO}^\parallel$ ($cm^{-1}$) | $\Gamma^\parallel$ ($cm^{-1}$) | $\varepsilon_\infty^\perp$ | $\omega_{TO}^\perp$ ($cm^{-1}$) | $\omega_{LO}^\perp$ ($cm^{-1}$) | $\Gamma^\perp$ ($cm^{-1}$) |
|---|---|---|---|---|---|---|---|---|
| $h^{10}B^{14}N$ [35] | 2.5 | 785 | 845 | 1 | 4.72 | 1392 | 1640 | 3.66 |
| $h^{10}B^{15}N$ (This work) | 2.99 | 774 | 829 | 2.68 | 4.38 | 1374 | 1624 | 4.05 |
| Natural hBN [31,35] | 2.95 | 760 | 825 | 3.0 | 4.90 | 1360 | 1614 | 7.00 |
| $h^{11}B^{14}N$ [35] | 3.15 | 755 | 814 | 1 | 4.62 | 1356 | 1610 | 3.36 |
| $h^{11}B^{15}N$ (This work) | 2.39 | 750 | 809 | 0.2 | 4.01 | 1339 | 1582 | 4.09 |

*Table 1: Extracted in-plane (⊥) and out-of-plane (∥) permittivity parameters of multicomponent hBN isotopes after fitting the polarization dependent reflectance data using the commercial software WVASE.*[44]

We next used scattering-type scanning near-field optical microscopy (s-SNOM) to image room-temperature HPhPs in upper RB of the doubly enriched hBN crystals. Fig. 2(a,b) present near-field amplitude maps of $h^{10}B^{15}N$ (at 1446 $cm^{-1}$) and $h^{11}B^{15}N$ (at 1406 $cm^{-1}$) flakes of comparable thickness on 300 nm $SiO_2$/Si substrates. Both maps reveal high-contrast polariton interference fringes parallel to the flake edge, consistent with contributions from edge-launched HPhPs and tip-launched HPhPs reflected by the edge. These fringes confirm the excitation and propagation of HPhPs in both materials. Differences are observed in both the fringe periodicity and spatial decay. The observed differences in fringe spacing and spatial decay reflect the combined influence of isotope composition, excitation frequency and the small variation in flake thickness. At their respective measurement conditions, the 51 nm thick $h^{10}B^{15}N$ crystal measured at 1446 $cm^{-1}$ supports a polariton wavelength of approximately 1580 nm, whereas the 57 nm thick

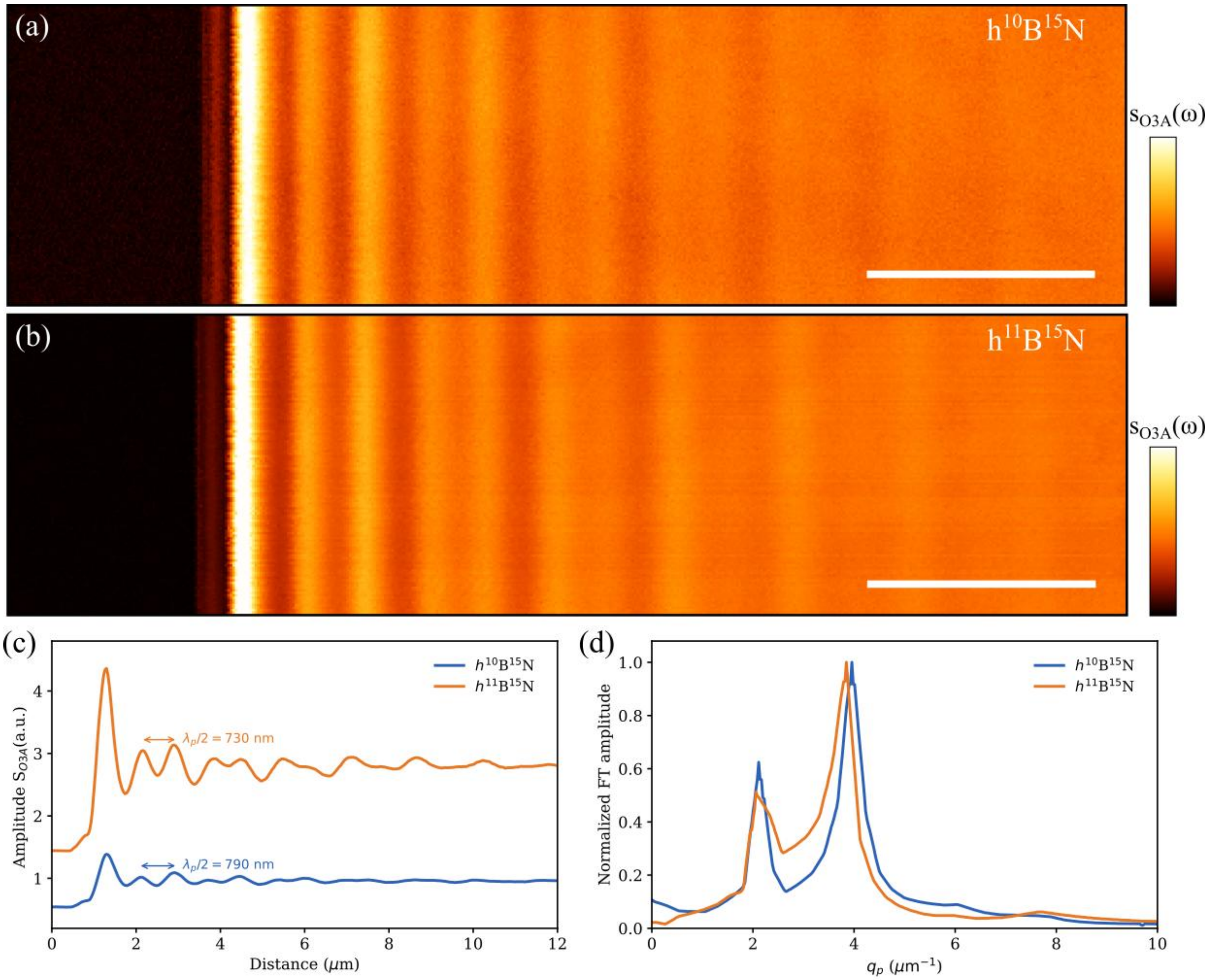

*Figure 2: s-SNOM imaging and line profile analysis of phonon polaritons in isotopically enriched hBN. Top panels: near-field third harmonic amplitude images of a (a) 51 nm thick $h^{10}B^{15}N$ crystal at 1446 cm$^{-1}$ and (b) 57 nm thick $h^{11}B^{15}N$ crystal 1406 cm$^{-1}$, showing polariton interference fringes parallel to the flake edges. (c) corresponding s-SNOM amplitude line profiles extracted from the near-field images. (d) Normalized Fourier-transform spectra of near-field line profiles for $h^{10}B^{15}N$ (blue) and $h^{11}B^{15}N$ (orange), showing the momentum-space profile of the propagating modes. The scale bar in near-field amplitude images is 3 μm.*

$h^{11}B^{15}N$ crystal measured at 1406 cm$^{-1}$ supports polariton wavelength of approximately 1460 nm, extracted from the line profile in Fig. 2(c). The lower boron mass in $h^{10}B^{15}N$ blue shifts the optical phonon frequencies, RBs, and the corresponding polariton dispersion relative to $h^{11}B^{15}N$, also evident from the FTIR. The slightly narrower phonon linewidths extracted from the far-field spectra are consistent with this trend, particularly for the out-of-plane mode (Table 1). The Fourier-transform spectra of the near-field line profiles, shown in Fig. 2(d), reveal the momentum-space content of the propagating modes at representative frequencies for $h^{10}B^{15}N$ and $h^{11}B^{15}N$, respectively. In both samples, the two principal Fourier peaks are consistent with edge-launched and tip-launched, edge reflected contributions to the fundamental HPhP mode, confirming that the fringes arise from guided slab modes rather than simple edge scattering.[45]

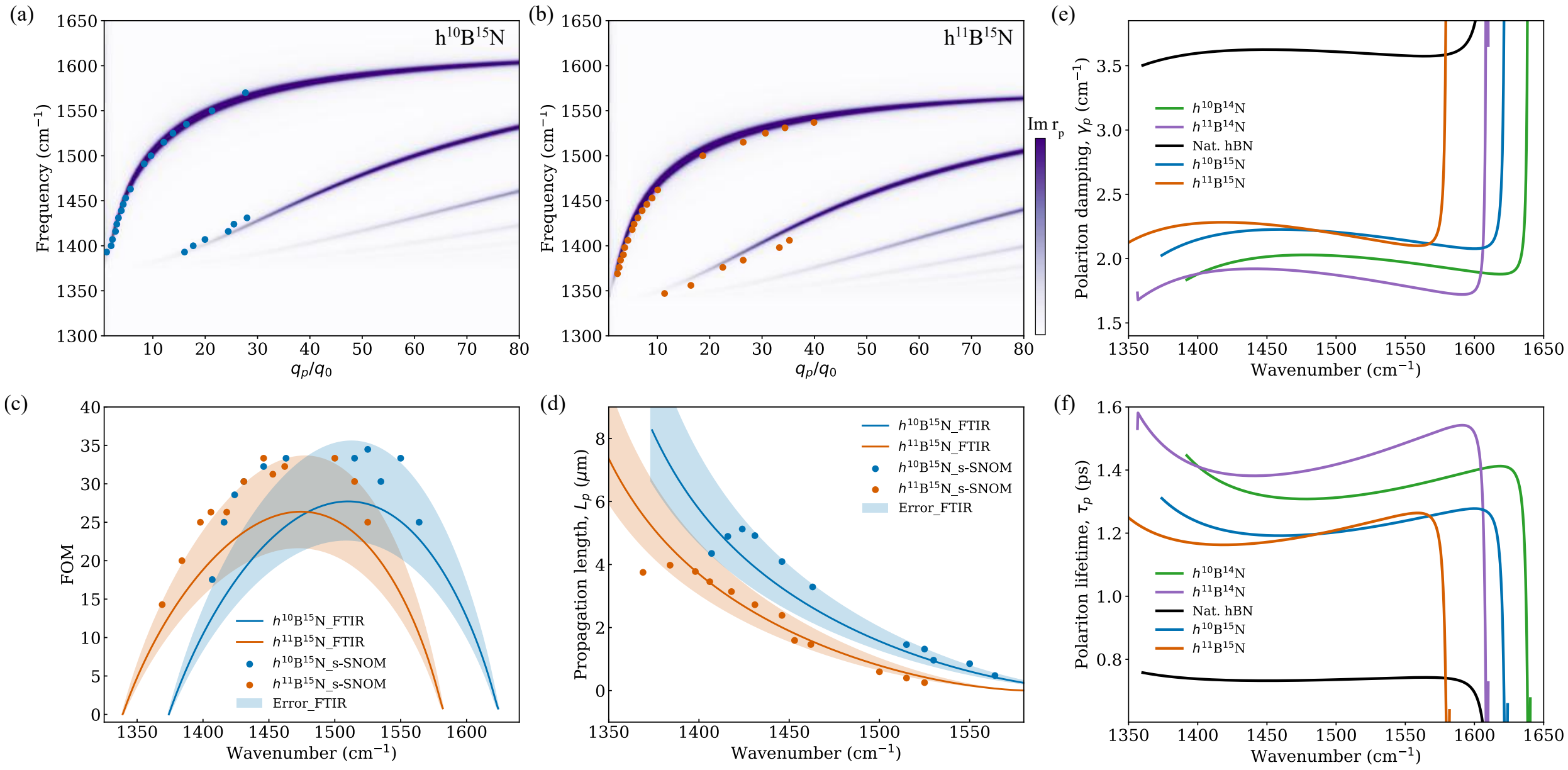

Figure 3*: Dispersion and loss characteristics of HPhPs in isotope engineered hBN. (a, b) Simulated HPhPs dispersion maps, represented by Im($r_p$), for 51 nm* h$^{10}$B$^{15}$N *and 57 nm* h$^{11}$B$^{15}$N *on 300 nm $SiO_2$/Si, respectively. Experimentally extracted polariton wavevectors are overlaid as symbols. (c) Propagating figure of merit, FOM and - 1/e intensity propagation length. $L_p$, for* h$^{10}$B$^{15}$N *and* h$^{11}$B$^{15}$N. *Symbols represent values extracted from the s-SNOM measurements, solid curves are calculated from the FTIR-derived dielectric functions, and shaded region show bounded sensitivity envelopes obtained by propagating ±1cm$^{-1}$ variation in the fitted TO, LO, and damping parameters. (e) Calculated polariton damping, $\gamma_p$ and (f) lifetime, $\tau_p$ for* h$^{10}$B$^{15}$N, h$^{11}$B$^{15}$N, *natural hBN,* h$^{10}$B$^{14}$N *and* h$^{11}$B$^{14}$N *respectively. All calculated quantities correspond to the upper RB HPhPs branch.*

To investigate the influence of isotopic mass on polariton momentum and dispersion, we analyzed how the guided HPhP mode structure, comprising the fundamental and higher order slab modes, evolves with slab thickness and composition. The experimentally extracted wavevectors were overlaid on the calculated dispersion maps obtained using the far field derived permittivity functions, as shown in Fig. 3(a,b). The experimental points (blue for h$^{10}$B$^{15}$N and orange for h$^{11}$B$^{15}$N) lie along the corresponding bright branches of the calculated spectral weight, indicating good agreement between near field experiment and the far field extracted dielectric model. The h$^{10}$B$^{15}$N dispersion shifts to higher frequency relative to h$^{11}$B$^{15}$N, consisting with its higher optical phonon frequencies. Within their overlapping upper RBs, h$^{11}$B$^{15}$N has the larger in-plane wavevector when compared at the same frequency, slab thickness and branch index. The additional branches in the calculated maps represent higher order guided HPhP modes. In the strongly confined regime, reducing the thickness increases the wavevector and confinement of a given branch, whereas increasing thickness brings higher order branches into a lower wavevector range. (see Supporting Information S3) Compared with previously reported single-isotope hBN benchmark[31,35], in which boron was enriched while nitrogen remained at natural abundance, the present doubly enriched crystal extend isotope control to both sublattices while retaining the characteristic guided HPhP band structure. These results demonstrate that multicomponent isotope engineering shifts the optical phonon frequencies and RBs, thereby tuning the frequency-wavevector dispersion of guided HPhPs.

Next, to quantify polariton loss, we determined the complex in-plane polariton wavevector, $q_p = q' + iq''$, and evaluated the propagating figure of merit, FOM = $Re(q_p)/Im(q_p)$, together with the 1/e intensity propagation length, $L_p = 1/[2Im(q_p)]$ (see Supporting Information S4).[31,36] The experimental values in Fig. 3(c,d) were obtained from Lorentzian fits to the geometrically corrected tip-launched polariton peaks in the Fourier-transformed s-SNOM line profiles. The solid curves were calculated independently from the FTIR-derived complex permittivity tensors and using the fundamental HPhPs branch. Because the calculated FOM and $L_p$ are sensitive to small variations in the fitted phonon parameters, we propagated a conservative ±1 $cm^{-1}$ variation in each in-plane and out-of-plane TO frequency, LO frequency, and damping constant. The shaded regions in Fig. 3(c,d) represent the bounded minimum-to-maximum sensitivity envelopes obtained from this six-parameter interval. (see Supporting Information S4) They therefore describe the sensitivity of the calculated quantities to the fitted optical parameters rather than conventional standard-deviation intervals. Both doubly enriched crystals exhibit broad FOM maxima within their respective upper RBs, and the experimental values follow the frequency dependence predicted from the far-field dielectric functions. The largest measured FOM values are approximately 34 for $h^{10}B^{15}N$ and 33 for $h^{11}B^{15}N$ with corresponding maximum measured propagation lengths of approximately 5.1 μm and 4.0 μm, respectively. These maxima correspond to the respective measurement frequencies and flake thicknesses. Overall, the agreement between the near-field measurements and the calculated curves demonstrates that the FTIR-derived dielectric functions capture both the dispersive and dissipative components of the polariton response. Fig. 3(e,f) further compare the calculated polariton damping and lifetime across the five hBN isotope compositions listed in Table 1. The polariton damping and lifetime evaluated as

$$\gamma_p = \frac{\omega |v_g|}{\text{FOM}\, v_p}, \qquad \tau_p = \frac{L_p}{|v_g|}$$

Where $v_g$ and $v_p$ are the polariton group and phase velocities, respectively.[35,38] (see Supporting Information S5) Away from the immediate RB edges, natural-abundance hBN exhibits $\gamma_p \approx 3.5$–3.7 $cm^{-1}$ and $\tau_p \approx 0.73$–0.76 ps. In contrast, the isotope-enriched compositions exhibit $\gamma_p \approx 1.7$–2.3 $cm^{-1}$ and $\tau_p \approx 1.2$–1.5 ps, corresponding to an approximately 1.6–2.1-fold lifetime enhancement relative to natural hBN. For the doubly enriched crystals investigated here, $\gamma_p$ remains approximately 2.0–2.3 $cm^{-1}$ and $\tau_p$ approximately 1.2–1.3 ps over most of the band interior. These crystals retain the low-loss response of isotopically purified hBN while shifting the RBs through control of both sublattice masses. Their polariton damping and lifetime remain comparable to monoisotopically enriched hBN because isotope-disorder scattering is strongly suppressed.[30]

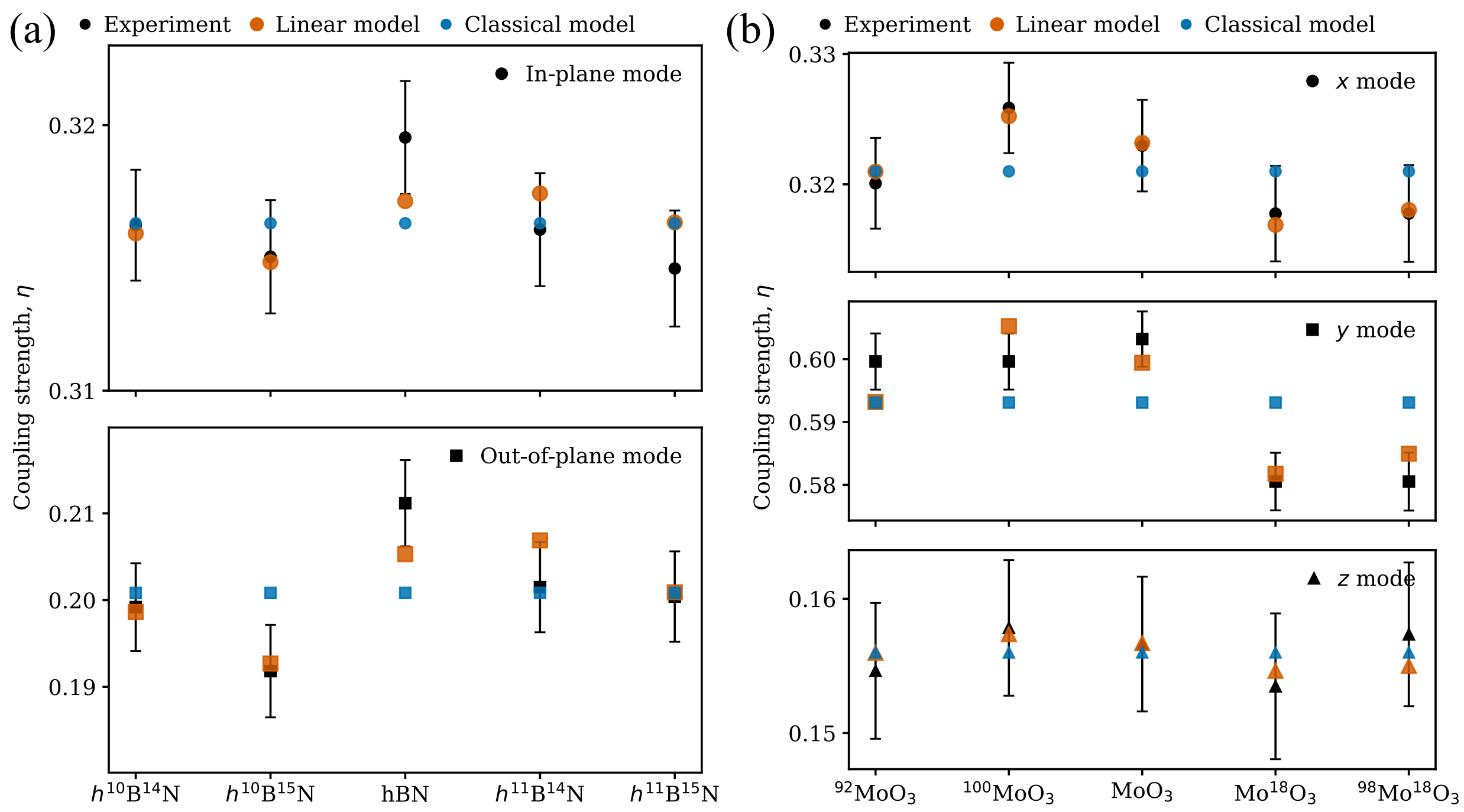

*Figure 4: Isotope dependence of the coupling strength in (a) hBN and (b) α-$MoO_3$. In hBN, circles and squares denote the in-plane and out-of-plane phonons, respectively. In α-$MoO_3$, circles, squares, and triangles denote the x, y, and z phonons, respectively. Black symbols are calculated directly from the measured TO and LO frequencies; their error bars are obtained by propagating an assigned 2 $cm^{-1}$ uncertainty for each frequency. Orange symbols are predictions of the two-component linear regression model, fit independently to the TO and LO data. Blue symbols are predictions of the conventional classical reduced-mass model.*

To determine whether the isotope dependence of the phonon response is adequately described by a single reduced mass, we analyzed the TO and LO frequencies separately for every isotope composition. In the conventional diatomic description, the optical-phonon frequency follows $\omega = C/\sqrt{\mu}$,[46–48] where $\mu = m_1m_2/(m_1 + m_2)$ and C is a mode-dependent constant. It assumes, however, that the two sublattices enter the frequency through the same reduced-mass combination. We therefore compared it with a two-component linear regression model, $\omega^2 = c_1/m_1 + c_2/m_2$, in which the coefficients quantify the fitted sensitivity of a given mode to each mass coordinate.[48] (Supporting Information S6) The two models make distinctly different predictions for the dimensionless LO-TO coupling strength η,[49]

$$\eta = \frac{g}{2\omega_{TO}} = \frac{\sqrt{\omega_{LO}^2 - \omega_{TO}^2}}{2\omega_{TO}}$$

For the classical reduced-mass model, the factor $\mu^{-1/2}$ cancels and η become almost isotope independent. In contrast, the two-component model retains the separate mass terms in both numerator and denominator and can therefore describe composition-dependent η. This distinction is tested directly in Fig. 4, where black symbols are calculated directly from the measured Optical Phonon (TO-LO frequencies), orange symbols are obtained from the two-component linear fit, and blue symbols are obtained from the classical reduced-mass fit. The deviations from the classical reduced-mass model are more pronounced in the individual TO and LO phonon frequencies. (Supporting Information S6 & S8)

For hBN (Fig. 4a), we compiled the previously reported isotope compositions $h^{10}B^{14}N$, $h^{11}B^{14}N$,[31,35,36,38] and natural abundance hBN [13] together with the nitrogen-enriched $h^{10}B^{15}N$ and $h^{11}B^{15}N$ crystals measured in this work. The reported TO and LO frequencies for nominally comparable hBN isotope compositions vary slightly among independent studies because they are extracted from separate experiments and dielectric-function fits. To account for this inter-study spread, while retaining a common uncertainty treatment for all compositions, we manually assign an uncertainty of 2 $cm^{-1}$ to each experimental TO and LO frequency.[35] This value is propagated through the calculation of $\eta$, producing the experimental error bars in Fig. 4. The error bars therefore represent the effect of the frequency uncertainty on the derived coupling strength. (Supporting Information S6) The hBN fits show that the two sublattice coefficients are of comparable magnitude: $c_B/c_N$ ranges from 0.946 to 1.361 across the four fitted phonons. (Fig S5 of Supporting Information S8) At the natural-composition reference point, the boron contribution to the fitted $\omega^2$ is approximately 55–64%, while the nitrogen contribution is approximately 36–45%. The generalized model consequently gives a modest improvement over the classical reduced-mass model, most clearly for the perpendicular modes. Because hBN lies close to the balanced reduced-mass limit, the resulting isotope dependence of $\eta$ is subtle and, for several compositions, comparable to the propagated uncertainty.

Next, we include $\alpha$-$MoO_3$ (Fig. 4b) as a comparison system that provides a stronger test of the analytical model. In hBN, the mass sensitivities of B and N are relatively balanced, so a conventional reduced-mass description remains a reasonable first approximation. $\alpha$-$MoO_3$, by contrast, allows the consequence of unequal sublattice mass sensitivity to be exposed directly. We analyzed the available isotope-dependent $\alpha$-$MoO_3$ phonon data for both Mo-enriched and O-enriched compositions, including $\alpha$-$^{92}MoO_3$, $\alpha$-$^{100}MoO_3$, natural $\alpha$-$MoO_3$, $\alpha$-$Mo^{18}O_3$, and $\alpha$-$^{98}Mo^{18}O_3$.[32,33,50–52] (Supporting Information S7) Here, the two-component fit reveals a strongly oxygen-dominated response. For the x and z phonons, $c_{Mo}/c_O$, total = 0.144–0.381 and 0.261–0.311, respectively; for y the ratio goes to 0.098 (Fig S6 of Supporting Information S8). Using the natural isotopic abundance masses of Mo and O to evaluate the two fitted inverse mass term, the oxygen contributes approximately 90% of the fitted $\omega^2$ phonon mode. The O-isotope substitution therefore produces the dominant phonon shift, whereas changing the Mo isotope alone produces only a small shift. The two-component model captures these O-driven changes in both the individual TO/LO frequencies and $\eta$ (Fig. 4b), while the one-parameter reduced-mass model cannot reproduce the strongly mode-selective behavior. The fitted coefficients represent effective mass sensitivities rather than microscopic force constants or phonon-eigenvector amplitudes. This framework resolves unequal sublattice contributions that are obscured by the conventional reduced-mass description and can be applied to isotope-dependent Raman and infrared-active phonons to predict shifts caused by substitution of each constituent.

## Conclusion

In summary, we used far- and near-field infrared measurements to determine how simultaneous boron and nitrogen isotope enrichment modifies the phonons modes and low loss HPhPs of hBN. At fixed $^{15}N$ composition, changing the boron isotope shifts the optical phonons, translates both RBs, and correspondingly shifts the HPhPs dispersion while retaining the guided HPhP response. Across the five-isotope composition, the combined frequency coverage extends from 750 to 845 $cm^{-1}$ for the lower RB and from 1339 to 1640 $cm^{-1}$ for the upper band, expanding the spectral range available beyond natural hBN alone. The doubly enriched crystals exhibit lower polariton damping and longer lifetimes than natural hBN, with values comparable to previously reported monoisotopically enriched crystals. The two-component inverse-mass analysis shows that the isotope response cannot always be represented by a single reduced mass, it resolves the contributions of the individual atomic masses and captures the strongly oxygen-dominated phonon shifts in α-$MoO_3$. Multicomponent isotope composition therefore provides a general materials design strategy by positioning the HPhPs spectral range while preserving low-loss propagation in polar van der Waals crystals.

## Methods

**Synthesis of hBN isotope:** Isotopically enriched $h^{10}B^{15}N$ and $h^{11}B^{15}N$ single crystals were synthesized using a two-step metal flux growth technique.[39] Isotopically pure boron (either $^{10}B$ or $^{11}B$) was alloyed with Ni and Cr in an argon-hydrogen atmosphere at 1550 °C to form a metal ingot. In a subsequent growth step, the ingot was exposed to sealed $^{15}N_2$ gas and slowly cooled to promote hBN crystal precipitation at the flux surface. The resulting crystals, were mechanically exfoliated for optical characterization.

**Far-field infrared reflectance spectroscopy.** Free-standing hBN isotope crystals were characterized by custom-built micro-FTIR microscope equipped with a Bruker Vertex 80V spectrometer. [35] Reflectance was measured in a co-polarized polarizer–analyzer configuration using a 40× Cassegrain objective (NA = 0.78) with an average incidence angle of ~38°. A custom aperture plate mounted above the microscope objective defined an restricted annular illumination sector, enabling controlled selection of the *p*- and *s*-polarized components incident on the sample and ensuring that the measured reflectance corresponded to a well-defined polarization state.[42] Spectra were collected over 500–6000 $cm^{-1}$ to capture the RBs of hBN and the higher-frequency permittivity required to constrain the dielectric-function fit. Low-temperature measurements were performed using an Oxford Instruments liquid-helium cryostat equipped with a ColdEdge Hydra closed-cycle chiller and ZnSe or KRS-5 windows, described previously in Ref.[35]

**Scattering-type scanning near-field optical mapping of hBN:** Mid-IR near-field imaging of nitrogen enriched isotope-engineered hBN was carried out using s-SNOM operating in tapping mode. Monochromatic mid-IR quantum cascade laser (QCL) radiation with excitation frequency ranging from 845 to 1800 $cm^{-1}$ (*p*-polarized illumination) was focused onto the Pt/Ir-coated silicon AFM tip (NanoAndMore, tip radius ~25 nm), and the near-field signal was recorded using pseudo-heterodyne interferometric detection with a tapping amplitude of ~70 nm. and an AFM tapping frequency of ~280 kHz. The detected near-field signal was demodulated at third harmonic of AFM tapping frequency to obtain background-free near-field images.

## Acknowledgement

N.R.S. and T.G.F. acknowledge funding support from the Multi-University Research Initiative (MURI) on Twist-Optics, sponsored by the Office of Naval Research under Grant No. N00014-23-1-2567. B.S. and T.G.F. also acknowledge support under NSF Grant No. 2318049, "Phonon Polariton Based Infrared Optoelectronics". H.Z acknowledges support under ONR Grant No. N00014-23-1-2616, "Ultrafast, nano-optic and temperature-dependent infrared (IR) probes for wide bandgap semiconductor characterization.". S.D. and B.N. acknowledge the support from the National Science Foundation under Grant No. DMR-2525882. E.J. and J.H.E. acknowledge support from the Office of Naval Research, award N00014-22-1-2582, for hBN crystal growth.

## Conflict of Interest

The authors declare no conflict of interest.

## Data Availability Statement

All data collected and analyzed in this study is available upon request to the corresponding authors.